# Analysis of the size of Solar system close to the state with zero total angular momentum via *Sundman's inequality*

*or*

*Solar system size analysis via Sundman inequality*

**Sergey Ershkov\***, Affiliation[1]: Plekhanov Russian University of Economics,

36 Stremyanny Lane, Moscow, 117997, Russia, Scopus number 60030998,

e-mail: sergej-ershkov@yandex.ru

Affiliation[2]: Sternberg Astronomical Institute, M.V. Lomonosov's Moscow State University, 13 Universitetskij prospect, Moscow 119992, Russia

**Dmytro Leshchenko**, Odessa State Academy of Civil Engineering

and Architecture, Odessa, Ukraine, e-mail: leshchenko_d@ukr.net

**Abstract**

In this paper, we present a new mathematical approach or solving procedure for analysis of the *Sundman's inequality* (for estimating *the moment of inertia* of the Solar system's configuration) with the help of *Lagrange-Jacobi relation*, under additional assumption of decreasing of the total angular momentum close to the zero absolute magnitude in the final state of Solar system in a future. By assuming such the final state for Solar system, we have estimated the mean-size of Solar system *R* via analysis of the *Sundman's inequality*.

So, to answer the question "Does the ninth planet exist in Solar system?", one should meet the two mandatory criteria for such the ninth planet, first is that it should have the negligible magnitude of inclination of its orbit with respect to the invariable plane. The second condition is that the orbit of the ninth planet should be located within the estimation for the mean-size of Solar system *R*.

## 1. Introduction, basic Sundman's inequality.

Let us consider the main features of general dynamics of Solar system (under the action of gravitational forces), which is governed by Sun and 8 celestial bodies of giant masses: in this problem $M_{Sun}$ and $m_i$ respectively, $i = \{1, \ldots, 8\}$, $m_i \ll M_{Sun}$, all the planets are rotating around their common centre of masses or barycenter on Kepler's trajectories (as first approximation). According to Arnold 1978, dynamics of Solar system could be considered in the Cartesian coordinate system $\vec{r} = \{x, y, z\}$ in case of Many Bodies Problem (*at given initial conditions*), with its origin, located at center of Sun, and coordinates of barycenter of the entire dynamical configuration

$$\vec{r}_G = \frac{1}{(m_i + M_{Sun})} \sum (m_i \cdot \vec{r}_i + M_{Sun} \cdot \vec{r}_{Sun})$$

whereas, *moment of inertia* of the Solar system's configuration should be defined as

$$I = \sum \left( m_i \cdot (\vec{r}_i - \vec{r}_G)^2 + M_{Sun} \cdot (\vec{r}_{Sun} - \vec{r}_G)^2 \right)$$

which is associated with the mean-size of Solar system $R$

$$R \cong \sqrt{\frac{I}{\sum (m_i + M_{Sun})}} \qquad (*)$$

Meanwhile, center of Sun is moving around the aforementioned barycenter on quasi-periodic trajectory (less than $2.19\, R_{Sun}$ from barycenter Jose 1965, where $R_{Sun}$ is the radius of Sun).

It is well-known from the theory of celestial mechanics Arnold 1978 the *Sundman's inequality*:

$$I \cdot K - J^2 \geq \left| \vec{C} \right|^2 \qquad (1)$$

with denotations

$$K = \sum \left(m_i \cdot (\dot{\vec{r}}_i - \dot{\vec{r}}_G)^2 + M_{Sun} \cdot (\dot{\vec{r}}_{Sun} - \dot{\vec{r}}_G)^2\right),$$

$$J = \sum \left(m_i \cdot ((\vec{r}_i - \vec{r}_G),(\dot{\vec{r}}_i - \dot{\vec{r}}_G)) + M_{Sun} \cdot ((\vec{r}_{Sun} - \vec{r}_G),(\dot{\vec{r}}_{Sun} - \dot{\vec{r}}_G))\right),$$

$$\vec{C} = \sum \left(m_i \cdot [(\vec{r}_i - \vec{r}_G) \times (\dot{\vec{r}}_i - \dot{\vec{r}}_G)] + M_{Sun} \cdot [(\vec{r}_{Sun} - \vec{r}_G) \times (\dot{\vec{r}}_{Sun} - \dot{\vec{r}}_G)]\right).$$

Here in the denotations above, $K$ means the kinetic energy of the entire dynamical configuration of Solar system; $J$ means the scalar product of the radius-vector of each object in the Solar system (including planets and Sun) onto appropriate velocity vector of the same object respectively; $\vec{C}$ denotes the pseudo-vector of total angular momentum of the system.

According to Chenciner 1997, equations of motions for the Solar system yield the *Lagrange-Jacobi relation*:

$$\dot{J} = \frac{\ddot{I}}{2} = K + 2kU, \qquad (2)$$

$$\dot{H} = 0$$

where $H = \frac{1}{2}K - U$ is the total energy of the system (sum of the kinetic energy $K$ in a Galilean frame fixing the center of mass and of the potential energy $U$), with $k = -\frac{1}{2}$ in the Newtonian case. Eliminating $K$ with the help of *Lagrange-Jacobi relation* (2), one transforms *Sundman's inequality* (1) into the differential inequality

$$I \cdot \ddot{I} - 2I \cdot H - \frac{1}{4}(\dot{I})^2 - \left|\vec{C}\right|^2 \geq 0 \qquad (3)$$

where $H = const$ if we consider Solar system as closed dynamical configuration with absence of interchange by the total energy with boundaries of this system along with zero level dissipation of energy inside (ejected in material of planets and gravitational waves or radiation leaving the Solar System, where thermal radiation is the primary means of energy transfer).

## 2. Inclination of orbits of planets in the Solar system, invariable plane.

The invariable plane of a planetary system (Laplace's invariable plane Souami & Souchay 2012) is defined as the plane passing through its barycenter perpendicular to its angular momentum pseudo-vector. In the Solar system, about 98% of the aforementioned total angular momentum is contributed by the orbital angular momentum of the four Gas giants (Jupiter, Saturn, Uranus and Neptune). It is worth to note that Jupiter's contribution is circa 60.3% to the Solar System's angular momentum, Saturn is circa at 24.5%, Neptune at 7.9%, and Uranus at 5.3%; meanwhile, the orbital angular momentum of the Sun and all non-jovian planets, minor planets, moons, and small Solar system's bodies, as well as the axial rotation momentum of all bodies, including the Sun, is in total only circa 2% of the Solar System's angular momentum (we should note that tidal phenomena in Solar system allowing the transfer of a infinitesimal amount of angular momentum from axial rotations to orbital revolutions due to tidal friction, and *vice-versa*).

As we can see from the Table 1, the invariable plane is within circa 1° of the orbital planes for all the 4 jovian planets:

Table 1. Inclinations of orbits of planets in the Solar system.

| Body | | Inclination to | | |
|---|---|---|---|---|
| | | Ecliptic | Sun's equator | Invariable plane |
| **Terrestrials** | Mercury | 7.01° | 3.38° | 6.34° |

|  | Venus | 3.39° | 3.86° | 2.19° |
|---|---|---|---|---|
|  | Earth | 0 | 7.155° | 1.57° |
|  | Mars | 1.85° | 5.65° | 1.67° |
| **Gas giants** | Jupiter | 1.31° | 6.09° | 0.32° |
|  | Saturn | 2.49° | 5.51° | 0.93° |
|  | Uranus | 0.77° | 6.48° | 1.02° |
|  | Neptune | 1.77° | 6.43° | 0.72° |
| **Minor planets** | Pluto | 17.14° | 11.88° | 15.55° |
|  | Moon | 5.09° (mean magnitude) | — | 6.59° |

The most extreme example of inclinations of Solar system's bodies is the Pluto's angle of orbit's inclination Ershkov & Leshchenko 2019 with respect to the invariable plane (which is 15°55'); namely, *Pluto's* orbit is *inclined*, or tilted, circa 15.5 degrees from the invariable plane. Except Mercury's inclination of 6°34' (and Moon's inclination of 6°59'), all the other orbits of planets in Table 1 are closer to invariable plane.

For the reason that the invariable plane is now within circa 1° of the orbital planes for all the 4 jovian planets (see Table 1), our main assumption would concern the future state of Solar system, close to the state with *zero total angular momentum*. Let us discuss this assumption at **Discussion** section below.

Indeed, if $\vec{C} \cong \vec{0}$, the motion takes place in a fixed plane (see seminal work Dziobek 1900).

### 3. Solving procedure for inequality (3) at zero total angular momentum.

Let us consider inequality (3). We should calculate the regions within Solar system or outside (close to its boundaries), where inequality (3) is valid under our main assumption $\vec{C} \cong \vec{0}$ (for the future state of Solar system, close to the state with *zero total angular momentum* of entire dynamical configuration, which is the natural restriction for the size of Solar system).

Aiming this way, let us transform inequality (3) to the form below:

$$I \cdot \ddot{I} - 2I \cdot H - \frac{1}{4}(\dot{I})^2 \geq 0 \quad \left\{ \left( \frac{dI}{dt} \right) = p(I) \rightarrow \left( \frac{d^2 I}{dt^2} \right) = \left( \frac{dp}{dI} \right) \cdot p \right\} \Rightarrow$$

$$I \cdot \frac{1}{2} \left( \frac{d(p^2)}{dI} \right) - 2I \cdot H - \frac{1}{4} p^2 \geq 0 \quad \left\{ p^2 = Y(I) \right\} \Rightarrow$$

$$\frac{dY(I)}{dI} - 4 \cdot H - \frac{1}{2} \left( \frac{Y(I)}{I} \right) \geq 0 \tag{4}$$

where $H = const$ should be determined or estimated additionally. Let us consider first the case of equality in (4) ($C_1 = const$):

$$\frac{dY(I)}{dI} - 4\cdot H - \frac{1}{2}\left(\frac{Y(I)}{I}\right) = 0 \qquad (5)$$

$$\Rightarrow$$

$$Y(I) = e^{-\int -\frac{1}{2}\left(\frac{dI}{I}\right)} \cdot \left[\int\left(4H\cdot e^{\int -\frac{1}{2}\left(\frac{dI}{I}\right)}\right)dI + C_1\right] = 8H\cdot I + C_1\cdot\sqrt{I}$$

Taking into account that $H > 0$, $I > 0$, $\sqrt{I} > 0$, $dI > 0$, and $Y(I) \geq 0$ in (4)-(5) (hence, we should choose $C_1 \geq 0$ in expression (5) for $Y(I)$), we obtain with the help of step-by-step transformations of inequality (6) below for $Y(I)$

$$Y(I) \geq 8H\cdot I + C_1\cdot\sqrt{I} \qquad (6)$$

which should be used in (4) to transform it to the form (6) again (besides, let us note that the process of integration at the inequality below should be by definition the proper summation of the chosen set of inequalities); these transfomations should prove the validity of inequality (6)

$$(4): \frac{dY(I)}{dI} - 4\cdot H \geq \frac{1}{2}\left(\frac{Y(I)}{I}\right), \quad (6): \frac{1}{2}\left(\frac{Y(I)}{I}\right) \geq 4H + C_1\cdot\frac{1}{2\sqrt{I}} \Rightarrow$$

$$(6)\to(4): \int dY \geq \int\left(8H + C_1\cdot\frac{1}{2\sqrt{I}}\right)dI \Rightarrow (6): Y(I) \geq 8H\cdot I + C_1\cdot\sqrt{I}$$

Thus, we obtain further from the differential inequality (6) as below

$$p^2 \geq 8H \cdot I + C_1 \cdot \sqrt{I} \quad \Rightarrow \quad \frac{dI}{dt} \geq \pm \sqrt{8H \cdot I + C_1 \cdot \sqrt{I}}$$

$$\Rightarrow \quad \int \frac{dI}{\sqrt{8H \cdot I + C_1 \cdot \sqrt{I}}} \geq \pm \int dt \quad \{\sqrt{I} = u\} \quad \Rightarrow \quad \int \frac{d\left(\sqrt{I} + \left(\frac{C_1}{8H}\right)\right)}{\sqrt{\sqrt{I} + \left(\frac{C_1}{8H}\right)}} \geq \pm \left(\sqrt{2H}\right) \cdot \int dt$$

$$\Rightarrow \quad \left\{ C_2 = \sqrt{\sqrt{I(t_0)} + \left(\frac{C_1}{8H}\right)} \geq 0 \right\} \quad \Rightarrow \quad \sqrt{\sqrt{I} + \left(\frac{C_1}{8H}\right)} - C_2 \geq \pm \left(\sqrt{\frac{H}{2}}\right) \cdot t \quad (7)$$

By using the Sundman inequality the aforementioned analysis (4)-(6) can be done, but still the end result (7) leaves open both possible solutions: a shrinking or expanding solar system. In case of Solar system's contraction in a future, we should choose sign "-" in inequality (7), it yields (all the physical magnitudes of constants in (7) are given in **Appendix A1**)

$$\left\{ \sqrt{\sqrt{I(t)} + \left(\frac{C_1}{8H}\right)} - \sqrt{\sqrt{I(t_0)} + \left(\frac{C_1}{8H}\right)} < 0 \right\} \quad \Rightarrow \quad \sqrt{\sqrt{I} + \left(\frac{C_1}{8H}\right)} \geq \sqrt{\sqrt{I(t_0)} + \left(\frac{C_1}{8H}\right)} - \left(\sqrt{\frac{H}{2}}\right) \cdot t$$

The last inequality means that Solar system should not be decreasing its size $R$ (*) less than

$$R \geq \frac{\left(\sqrt{R(t_0) \cdot \sqrt{\sum(m_i + M_{Sun})} + \left(\frac{C_1}{8H}\right)} - \left(\sqrt{\frac{H}{2}}\right) \cdot t\right)^2 - \left(\frac{C_1}{8H}\right)}{\sqrt{\sum(m_i + M_{Sun})}} \quad (8)$$

Meanwhile, the key parameter in estimation of size *R* in inequality (8) is the total energy *H* of Solar system.

Let us schematically imagine at Figs.1-2 the lower bound of the estimation for the Solar system's size *R* (depending on time *t*), which corresponds to the aforementioned inequality (8) (at Figs.1-2 we designate *x* = *t* just for the aim of presenting the plot of solution)

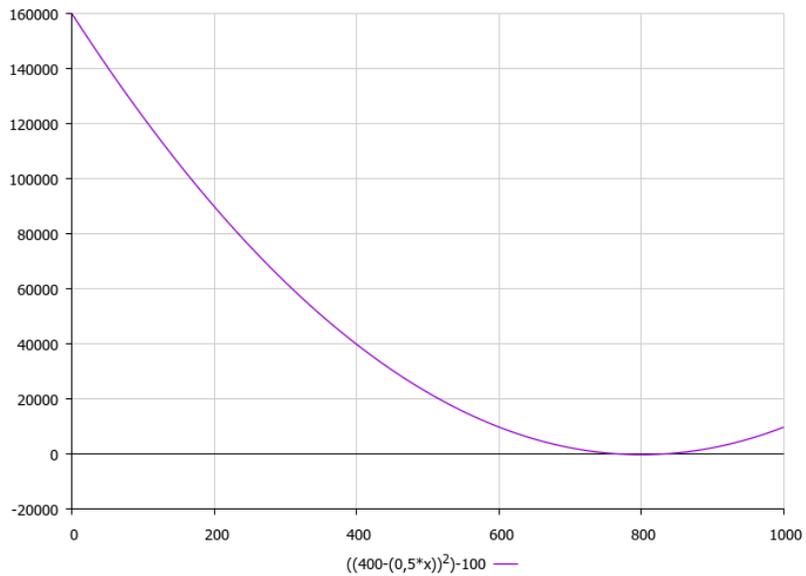

Fig.1. A *schematic* plot of the lower bound of the estimation

for the Solar system's size *R*(*t*) (8) (depending on time *t*).

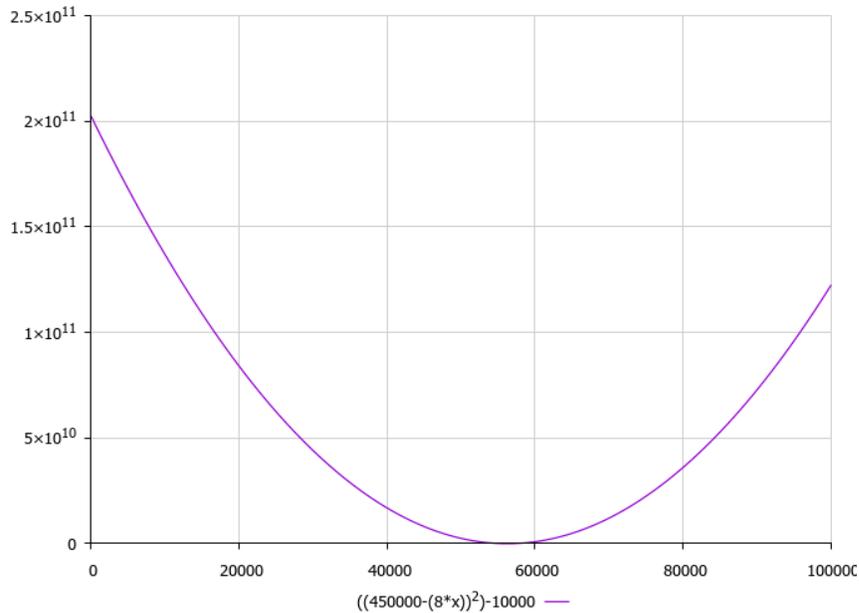

Fig.2. A *schematic* plot of the lower bound of the estimation for the Solar system's size $R(t)$ (8) (depending on time $t$).

The last but important note: if we choose the decreasing of Solar system's size during the evolution in a future as expected scenario, its magnitude should allow to exceed the level of minimal distances or Roche's limits Arnold 1978 for all the planets and Sun in our analysis (to avoid irreversible collisions of celestial bodies in Solar system).

**4. Discussion.**

As we can see, about 98% of the aforementioned total angular momentum is contributed by the orbital angular momentum of the four Gas giants (Jupiter, Saturn, Uranus and Neptune). Nevertheless, the total angular momentum of the system is not conserved: obviously, each of four Gas giants have been tidally interacting Ershkov 2017a with the Sun during a billions of years of their evolution in Solar system, allowing transfer of angular momentum in the infinitesimal amounts from their orbital revolutions to the axial rotation of Sun (due to tidal friction). But Sun is known to have been lost more than 90% of its angular momentum of axial rotation, due to the ejecting of angular momentum with solar wind (or solar plasma) outside the Solar system Kitchatinov 2005, 2006 (obviously, both the aforementioned processes are synchronized). Combined together, these are obviously the fundamental governing processes of the step-by-step decreasing of the total angular momentum to zero for most of the planets in Solar system - first of all, for the four Gas giants.

Nevertheless, there is a natural mechanism, preventing the aforesaid process of changing the angular momentum (or decreasing its absolute magnitude) to zero for planets in Solar system. This is the net transfer of angular momentum between planets which are locked into the resonance (due to dynamical features of the orbital evolution of planets under the influence of the resonance phenomenon). But, there are only a few known perfect mean-motion resonances in the Solar system, involving planets, dwarf planets or larger satellites:

1) 2:3 Pluto-Neptune;
2) 2:4 Tethys-Mimas (Saturn's moons), not classical case;
3) 1:2 Dione-Enceladus (Saturn's moons);
4) 3:4 Hyperion-Titan (Saturn's moons);
5) 1:2:4 Ganymede-Europa-Io (Jupiter's moons, ratio of orbits).

Meanwhile, the Pluto's angle of orbit's inclination with respect to the invariable plane (which is 15°55') could be explained by the presence of the aforementioned perfect mean-motion resonance 2:3 Pluto-Neptune (obviously, there is the net transfer of angular momentum between these planets).

Also we should note that there exist "a near resonance" phenomena for planets in Solar system. The presence of a near resonance may reflect that a perfect resonance existed in the past, or that the system is evolving towards one in the future Bazsó, *et al.* 2010.

Table 2. Most interesting "a near resonance" phenomena in the Solar system.

| (Ratio) and Bodies | Mismatch after one cycle | Possible period for matching | Probability |
|---|---|---|---|
| *Planets* | | | |
| (9:23) Venus-Mercury | 4.0° | 200 years | 0.19 |
| (8:13) Earth-Venus | 1.5° | 1000 years | 0.065 |
| (243:395) Earth-Venus | 0.8° | 50,000 years | 0.68 |

As we can see from Table 2, it is very likely a perfect resonance existed in the past between Earth and Venus (with the help of which transfer of angular momentum between these planets may have been initiating, that's why both the inclinations of Earth and Venus have been tilted additionally with respect to invariable plan, see Table 1). The same mechanism was possible for interaction between Venus and Mercury in the past which caused increasing of inclination of Mercury's orbit with respect to variable plane drastically, in total amount of 6°34'.

5.  **Conclusion.**

In this paper, we present a new mathematical approach or solving procedure for analysis of the *Sundman's inequality* (for estimating *the moment of inertia* of the Solar system's configuration) with the help of *Lagrange-Jacobi relation*, under additional assumption of decreasing of the total angular momentum close to the zero absolute magnitude in the final state of Solar system in a future.

The main reason for such assumption is that the invariable plane is now within circa 1° of the orbital planes for all the 4 jovian planets (see Table 1), whereas about 98% of the aforementioned total angular momentum is contributed by the orbital angular momentum of the four Gas giants (Jupiter, Saturn, Uranus and Neptune).

The basis for the aforesaid assumption one can find in the profound work of Dziobek 1900: if components of total angular momentum is zero, the motion takes place in a fixed plane. By assuming such the final state for Solar system, we have estimated the mean-size of Solar system $R$ (*) via analysis of the *Sundman's inequality*.

As the main findings, we have found also that the inclinations of orbits of planets in the Solar system will have been tidally decreasing forever to final state close to the zero angular momentum of entire dynamical configuration in a future.

So, to answer the question "Does the ninth planet exist in Solar system?" Batygin, *et al.* 2019, one should meet the two mandatory criteria for such the ninth planet, first is that it should have the negligible magnitude of inclination of its orbit with respect to the invariable plane (in case of absence the perfect mean-motion resonance, which can be definitely excluded for the Gas giants of Solar system). The second condition is that the orbit of the ninth planet should be located within the aforementioned estimation (7)-(8) for the mean-size of Solar system $R$ (*) (obtained via analysis of the *Sundman's inequality*).

The last but not least, we should especially note that at using of *Lagrange--Jacobi relation*, there is ambiguity in estimation of level of $H = const$, i.e. the total energy of the system (sum of the kinetic energy $K$ in the chosen Galilean reference frame and of the potential energy $U$). Indeed, Solar system is not a closed dynamical configuration with absence of interchange by the total energy outside its boundaries (e.g., gravitational waves or radiation leaving the Solar System, where thermal radiation is the primary means of energy lost).

Also, some remarkable articles should be cited, which concern the problem under consideration, Bruns 1887, Chernousko, *et al.* 2017, Duboshin 1968, Dvorak & Lhotka 2013, Ershkov 2015, Ershkov 2017b, Poincaré 1967.

**Appendix, A1 (useful data for estimating constants in inequality (7)).**

Let us write out the appropriate useful data for estimating the constants in inequalities (7)-(8) and formula (*) for the mean-size of Solar system *R*:

1) $\sum (m_i + M_{Sun}) = 1.014 \, M_{Sun} = 2.016 \cdot 10^{30} kg$

2) Taking into account the well-known formula for gravitational binding energy Burša, *et al.* 1996 $U_i = -\frac{3}{5}\frac{G \cdot m_i^2}{R_i}$, where $R_i$ is the radius of planet with mass *m i* (meanwhile, the kinetic energy of the planet $m_i$ circa ~ twice less than the absolute magnitude of the potential energy $U_i$), we can conclude that potential energy of Sun plays *a crucial role* in estimation for the magnitude of the total energy *H* of Solar system

$$U_{sun} = -\frac{3}{5}\frac{G \cdot M_{sun}^2}{R_{sun}} = -\frac{3}{5} \cdot \frac{6.6742 \cdot 10^{-11}[m^3 \cdot kg^{-1} \cdot s^{-2}] \cdot (1.98844 \cdot 10^{30} kg)^2}{6.957 \cdot 10^8 m}$$

$$\Rightarrow \qquad H \cong |U_{sun}| \cong 2.2759 \cdot 10^{41}[kg \cdot m^2 / s^2] \qquad (9)$$

3) As for the *current* absolute magnitude of the total angular momentum of the Solar system Cang, *et al.* 2016, we can estimate it as $|\vec{C}| = 3.3212 \cdot 10^{45} \, [kg \cdot m^2 / s]$

4) Taking into account the current age of the Solar system (*t* ~ 4.6 billions of years), let us assume that by virtue of comparing the right part of inequality

(7) to the constant $C_2$ (as well as comparing constant $C_1$ in combination with respect to other terms in the left part $\sqrt{\sqrt{I}+\left(\dfrac{C_1}{8H}\right)}$ ), the aforesaid constants could be chosen as below:

$$C_2 \sim \{1,...,9\}\cdot 10^{37}\,[kg^{\frac{1}{2}}\cdot m]\,,$$

asfor its mutual commensurability with other terms at both the parts of inequality (7), and

$$|C_1| \sim 1.17\cdot 10^{70}\,[kg^{\frac{3}{2}}\cdot m^3/s^2]$$

## Acknowledgments

Authors are thankful to unknown esteemed Reviewers with respect to their valuable efforts and advices which have improved structure of the article significantly.

## Conflict of interest

Authors declare that there is no conflict of interests regarding publication of article.

Remark regarding contributions of authors as below:

In this research, Dr. Sergey Ershkov is responsible for the general ansatz and the solving procedure, simple algebra manipulations, calculations, results of the article in Sections 1-4, **Appendix A1**, and also is responsible for the search of approximated solutions.

Prof. Dmytro Leshchenko is responsible for theoretical investigations as well as for the deep survey in literature on the problem under consideration.

All authors agreed with results and conclusions of each other in Sections 1-5.

**References**:

Arnold, V. Mathematical Methods of Classical Mechanics. Springer, New York, 1978.

Bazsó, A., Eybl, V., Dvorak, R., Pilat-Lohinger, E., and Lhotka, C. A survey of near-mean-motion resonances between Venus and Earth. Celestial Mechanics and Dynamical Astronomy. 107 (1): 63–76, 2010.

Batygin, K., Adams, F.C., Brown, M.E., and Becker, J.C. The planet nine hypothesis. Physics Reports, Vol. 805, 3 May 2019, pp. 1-53, 2019.

Bruns, H. Ueber die Integrale der Vielkoerper-Problems. Acta math. Bd. 11, p. 25-96, 1887.

Burša M., Křivský L., and Hovorková O. Gravitational potential energy of the Sun. Studia Geophysica et Geodaetica. Vol.40, Issue 1, pp.1-8, 1996.

Cang R., Guo J., Hu J., and He C. The Angular Momentum of the Solar System. Astronomy & Astrophysics. Vol. 4, Issue 2, pp. 33-40, 2016.

Chenciner, A. Introduction to the N-body problem (école d'été de Ravello, september 1997). See also at https://perso.imcce.fr/alain-chenciner/preprint.html, as the preprint, published in the year 1997:

https://perso.imcce.fr/alain-chenciner/1997Ravello.pdf

Chernousko, F.L., Akulenko, L.D., and Leshchenko, D.D. Evolution of motions of a rigid body about its center of mass. Springer, Cham, 2017.

Duboshin, G.N. Nebesnaja mehanika. Osnovnye zadachi i metody. Moscow: "Nauka" (handbook for Celestial Mechanics, in russian), 1968.

Dvorak, R., and Lhotka, Ch. Celestial Dynamics. Chaoticity and Dynamics of Celestial Systems. Wiley-VCH Verlag GmbH & Co. KGaA,Weinheim, Germany, 2013.

Dziobek, O. Ueber einen merkwürdigen Fall des Vielkörperproblems. Astron. Nacht. 152, p.33-46, 1900.

Ershkov, S.V. Stability of the Moons Orbits in Solar System in the Restricted Three-Body Problem. Advances in Astronomy, article number: 7, 2015.

Ershkov, S.V. About tidal evolution of quasi-periodic orbits of satellites. Earth, Moon and Planets. Vol. 120, №1, pp.15-30, 2017.

Ershkov, S.V. Forbidden Zones for Circular Regular Orbits of the Moons in Solar System, R3BP. Journal of Astrophysics and Astronomy.Vol. 38, no.1, pp.1-4, 2017.

Ershkov, S.V., and Leshchenko, D. Solving procedure for 3D motions near libration points in CR3BP. Astrophysics and Space Science. Vol.364, no.207, 2019.

Jose, P.D. Sun's motion and sunspots. Astron. J. Vol.70. №3. P.193-200, 1965.

Kitchatinov, L.L. The differential rotation of stars. Physics – Uspekhi. Vol.48, № 5. pp.449-467, 2005.

Kitchatinov, L.L. Differential rotation of a star induced by meridional circulation. Astronomy Reports. Vol.50, №6. p.512, 2006.

Poincaré, H. New Methods of Celestial Mechanics, 3 vols. (English trans.), 1967. American Institute of Physics.

Souami, D., and Souchay, J. The solar system's invariable plane. Astronomy and Astrophysics, 543: A133, 2012.